\documentclass[conference]{IEEEtran}

\usepackage{amssymb}
\usepackage{fancyhdr}
\usepackage[numbers,square,sort&compress]{natbib}

\usepackage{cite}
\usepackage[cmex10]{amsmath}
\usepackage{algorithm}
\usepackage{algpseudocode}
\usepackage{float,amsthm}
\usepackage{mdwmath}
\usepackage{mdwtab}
\usepackage[tight,footnotesize]{subfigure}
\usepackage[pdftex]{graphicx}
\usepackage{array}
\usepackage{booktabs}
\usepackage{multirow}
\usepackage{flushend}
\usepackage{amsxtra}

\theoremstyle{definition}

\newtheorem{definition}{Definition}

\newcounter{remark}

\newcounter{example}
\newenvironment{example}[1][Example]{\refstepcounter{example}\begin{trivlist}
\item[\hskip \labelsep {\bfseries #1 \theexample.}]}{\end{trivlist}}

\newcounter{property}

\begin{document}

\title{Measuring Conflict in a Multi-Source Environment as a Normal Measure}

\author{
\IEEEauthorblockN{Pan Wei$^1$,~\IEEEmembership{Student Member,~IEEE,}
John E. Ball$^1$,~\IEEEmembership{Senior Member,~IEEE,}
Derek T. Anderson$^1$,~\IEEEmembership{Senior Member,~IEEE,}
Archit Harsh$^1$,~\IEEEmembership{Student Member,~IEEE,}
and Christopher Archibald$^2$}\\
\IEEEauthorblockA{$^1$Department of Electrical and Computer Engineering, Mississippi State University\\
$^2$Department of Computer Science and Engineering, Mississippi State University\\
Mississippi State, MS 39762, USA\\
\{pw541@, jeball@ece., anderson@ece., ah2478@, archibald@cse.\}msstate.edu}
}

\maketitle



\begin{abstract}
In a multi-source environment, each source has its own credibility.
If there is no external knowledge about credibility then we can use the information provided by the sources to assess their credibility.
In this paper, we propose a way to measure conflict in a multi-source environment as a normal measure.
We examine our algorithm using three simulated examples of increasing conflict and one experimental example.
The results demonstrate that the proposed measure can represent conflict in a meaningful way similar to what a human might expect and from it we can identify conflict within our sources.
\end{abstract}

\IEEEpeerreviewmaketitle


\section{Introduction}\label{sec:Introduction}
Living in an imperfect world, we can never fully trust a source due to various reasons such as noise, faulty sensors, deception, etc.
In some scenarios we do not have external knowledge about the credibility of each source, the only information we have is the information from the sources.
Herein, we focus on measuring conflict from source information to help assess credibility.
In many applications, supposing each source is independent, we can use such a measure to help identify conflict within our sources.
In order to facilitate better decisions, we can put more trust on those sources who have less conflict and diminish the influence of the sources who have conflict above some amount.
However, we do note that in some cases conflicting cases can be the most interesting and deserve further analysis.

Herein, we focus on a new simple method to calculate measure of conflict.
Some theories exist to measure conflict: Shannon entropy \citep{baraka2014shannon}, fuzzy measure \citep{giordano2010fuzzy} and belief theory \citep{martin2008belief}.
On the other hand, a number of fuzzy measures of agreement have been put forth to date \citep{gower1971general}\citep{zezula2006similarity}\citep{wagner2012fuzzy}\citep{havens2013fuzzy}\citep{havens2014data}.
While related, conflict and agreement can be quite different and difficult to describe in terms of one another.
Conflict is a challenge to capture in terms of an algorithm that is \lq\lq in line\rq\rq\, with what a human expects.

Section II is background of normal measure and fuzzy measure.
Section III proposes the algorithm for calculating the measure of conflict (CF) and Section IV provides four numeric examples.
Section V is conclusion and future work.

\section{Background}\label{sec:Background}
To better understand the proposed measure of conflict, we first need to review the following definitions of a normal measure (NM) and a fuzzy measure (FM) \citep{garmendia2005evolution}\citep{grabisch2000fuzzy}.

Let $X=\{x_1,...,x_n\}$ be a set of $n$ information sources (e.g., experts, sensors, algorithms, etc.).

    \begin{definition}
    \textbf{(Normal measure)}
    \citep{garmendia2005evolution}
    Let (X, $\Omega$) be a measurable space, where $X$ is a set and $\Omega$ is a $\sigma$-algebra of $X$. A measure $g$: $\Omega$ $\to$ [0, 1] is a NM if there exists a minimal set $A_0$ (e.g., $\emptyset$) and a maximal set $A_m$ (e.g., $X$) in $\Omega$ such that:
     \begin{enumerate}
     \item $g(A_0) = 0$,
     \item  $g(A_m) = 1$.
     \end{enumerate}
    \end{definition}

    \begin{definition}
    \textbf{(Fuzzy measure)}
    \citep{grabisch2000fuzzy}\citep{garmendia2005evolution}
    Let (X, $\Omega$) be a measurable space. A measure $g$: $\Omega$ $\to$ [0,1] is a FM if it has the following properties:
      \begin{enumerate}
      \item (Normality) $g(\emptyset) = 0$ (and often $g(X)=1$),
      \item  If $A, B \in \Omega$ and $A \subseteq B \subseteq X$, then $g(A) \le g(B) \le 1$.
      \end{enumerate}
    \end{definition}

\begin{figure}[t!]
    \centering
     \subfigure{\includegraphics[height=.15\textheight]{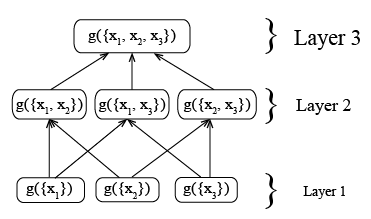}}
     \caption{Fuzzy measure lattice for three information sources.}
     \label{fig:layer}
    \end{figure}

\noindent Note, often $g(X)=1$ for problems like confidence/decision fusion; however the interval can and has been extended to domains like $[0,\Re^+]$. The difference between a FM and NM is that FM is monotone.
For example, if there are three sources ($x_1$, $x_2$, $x_3$) then the FM $g(\{x_1, x_2, x_3\}) \geq max\left(g(\{x_1, x_2\}), g(\{x_1, x_3\}), g(\{x_2, x_3\})\right)$, and $g(\{x_1, x_2\}) \geq max(g(\{x_1\}), g(\{x_2\}))$. The FM lattice for three sources is shown in Fig. ~\ref{fig:layer}. The measures calculated using the algorithm proposed in this paper is a NM. The reason for using a NM is that if we used a FM then adding a conflicting source could lessen our measure value; however, it cannot due to FM monotonicity.

\section{Measure of Conflict}\label{sec:MC}
In order to measure and identify conflict, we propose an algorithm called the measure of conflict (CF).
The CF is defined here as the sum of sub-interval conflicts over the all sources' measurement range. Each sub-interval contributes to the conflict if some of the sources' measurement range does not cover such sub-interval. The conflict metric is the ratio between the sub-interval length and the global measurement range length multiplied by the percentage of sources not covering that sub-interval.
The nomenclature and mathematical description of the algorithm is as follows:

Let $X = \{x_1, ...,x_n\}$ be a set of sources for which we do not know the credibility of each source.
Each source $x_i$ provides interval-valued evidence ($\overline h_i$).
For example, if sensor $x_1$ has output voltages $v$, where $2.0 \leq v \leq 3.5$, then the interval is $\overline h_1 = [2.0, 3.5]$, and this is called interval-valued evidence.
$A_i $ is the set that contains all i-tuple combinations of the sources, where $i$ is the number of sources in the tuple. For example, with sources $x_1, x_2$ and $x_3$, $ A_1 = \{\{x_1\}, \{x_2\}, \{x_3\}\}$, $ A_2 = \{\{x_1, x_2\}, \{x_1, x_3\}, \{x_2, x_3\}\}$, and $A_3 = \{x_1, x_2, x_3\} = X$.
$g$ is the numeric NM.

\begin{figure}[t!]
    \centering
     \subfigure{\includegraphics[height=.1\textheight]{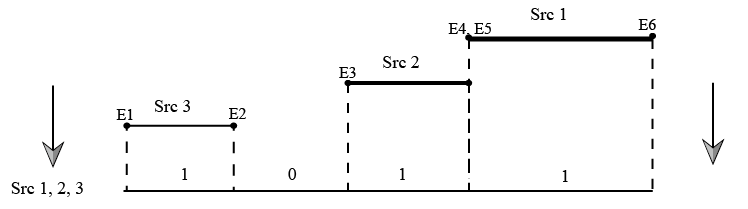}}
     \caption{Counting the number of sources with one region that has no sources and is marked as 0.}
     \label{fig:zerooverlap}
    \end{figure}

Suppose there are $n$ interval-valued sources. Let $E = \{E_1, ..., E_{2n}\}$ be the ordered set $(E_i\leq E_{i + 1})$ of all interval endpoints from our evidences. An example is shown in Fig. ~\ref{fig:zerooverlap}.
Let $\overline P = \{ \overline P_1, ..., \overline P_{2n - 1} \}$ be the set of intervals induced by $E$. For example, $\overline P_1 = [E_1, E_2]$. Let $O(\overline P_k)$ be the number of sources (which are in $A'\in A_i$) in interval $\overline P_k$. Let $g^{CF}$ be defined as:
\begin{subequations}
    \begin{equation}
     g^{CF}( A' \in A_1) = 0, \forall A' \in A_1
    \end{equation}

    \begin{multline}
     \tilde g^{CF}(A' \in A_i) =\sum \limits_{k=1}^{2n - 1} \Phi (\overline P_k, A') |\overline P_k| \left( \frac{i - O(\overline P_k)}{i}\right),\\
     \Phi (\overline P_k, A') =  \left\{
     \begin{array}{l}
   1 \quad  if \; \overline P_k \subseteq [ \min\limits_{x_j \in A'}  [ \overline h_{j}]^-, \; \max\limits_{x_j \in A'} [ \overline h_{j}]^+],\\
   0 \quad else,\\
   \end{array}
   \right. \\
   i = [2 : n],
    \end{multline}

    \begin{multline}
     g^{CF}(A_i) =
     \tilde g^{CF}(A_i) / (\max\limits_{x_j \in A'} [ \overline h_{j}]^+ - \min\limits_{x_j \in A'}  [ \overline h_{j}]^-).
      \end{multline}
\label{eqn:MC1.1}
\end{subequations}

\begin{figure}[t!]
    \centering
    \subfigure[Two different cases of three interval-valued sources where Case 1 has more conflict than Case 2. ]{\includegraphics[height=.11\textheight]{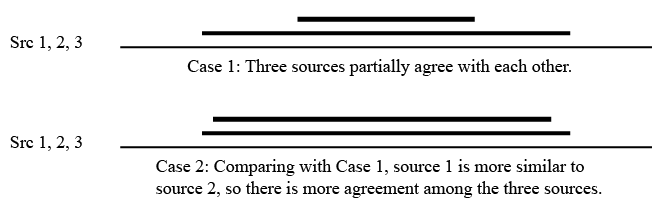}}
    \subfigure[Two different cases of two interval-valued sources where Case 2 has more conflict than Case 1. ]{\includegraphics[height=.11\textheight]{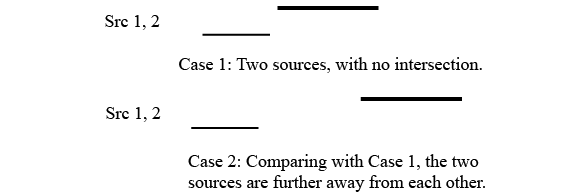}}
     \caption{Comparison of different cases in which one case has more conflict than the other case.}
     \label{fig:limitation}
    \end{figure}

\noindent In CF, every interval that is bounded within the maximum right and minimum left endpoints have been weighted based on the number of overlap times.
If an interval has no sources in it then it has the highest conflict weight of one.
On the other hand, if all sources overlap in an interval then there is no conflict and a weight of zero is assigned.
The CF not only considers intervals that have overlapping sources, but also considers all intervals without any overlapping sources (which is the section marked as 0 in Fig. ~\ref{fig:zerooverlap}).
Using CF, the differences between case 1 and case 2 in Fig. ~\ref{fig:limitation} (a) and (b) can be clearly shown. In Fig. ~\ref{fig:limitation} (a), comparing the two cases, there is more conflict among the three sources of case 1 than that of case 2 since the interval from source 1 is more similar to source 2 and therefore there is a wider overlapping region among the three sources. In Fig. ~\ref{fig:limitation} (b), case 2 has more conflict between the two sources since there is a wider no overlapping region versus case 1. These two examples show that CF works \lq\lq in line\rq\rq \,with what a human would expect.

\section{Numeric Examples}\label{sec:examples}

   \subsection{Synthetic Examples}
   In this section, three synthetic examples are provided to demonstrate the calculation and meaning of CF for increasing degrees of conflict.

   \begin{example}
   In the first example, there are four interval-valued evidences, $\overline h_1 = [0, 12]$, $\overline h_2 = [0, 4]$, $\overline h_3 = [0, 3]$, and $\overline h_4= [0, 2]$. The four evidences are shown in Fig ~\ref{fig:eg1}(a), while the lattice of CF conflict measures is shown in Fig ~\ref{fig:eg1}(b). The following is a manual calculation example using CF.

   \begin{figure}[t!]
    \begin{center}
    \subfigure[Four interval-valued sources.]{\includegraphics[height=.085\textheight]{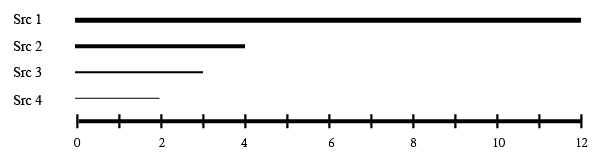}}
    \subfigure[Lattice of CF values.]{\includegraphics[height=.17\textheight]{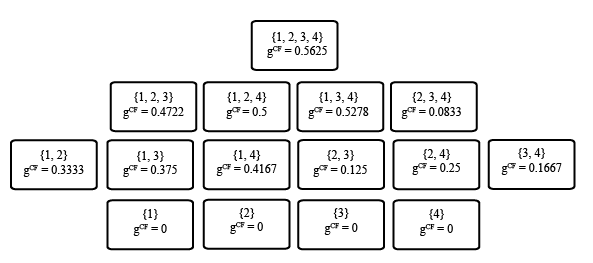}}
    \caption{Example 1: Small conflict.}
    \label{fig:eg1}
    \end{center}
    \end{figure}

    For the calculation of $g^{CF}(\{x_1, x_2, x_3\})$, as shown in Fig. ~\ref{fig:calculation}, interval [4, 12] has one source, which is source 1. Interval [3, 4] has two overlapping sources which are from source 2 and 3. Interval [0, 3] has all three overlapping sources. We can compute CF using Eq.~\ref{eqn:MC1.1} as follows:
~\\[6pt]
     $\tilde g^{CF}(\{x_1, x_2, x_3\}) = |12 - 4|\times \frac{3 - 1}{3} + |4 - 3|\times \frac{3 - 2}{3} = \frac {17}{3},$
~\\[6pt]
     $g^{CF}(\{x_1, x_2, x_3\}) = \frac {17}{3}  /  |12 - 0| \approx 0.4722$.
~\\[6pt]

    \begin{figure}[t!]
    \centering
     \subfigure{\includegraphics[height=.09\textheight]{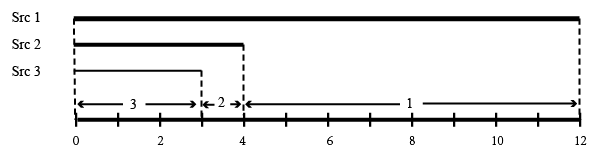}}
     \caption{An example showing how to calculate $g^{CF}{(\{x_1, x_2, x_3\})}$ of three sources.}
     \label{fig:calculation}
    \end{figure}
The results in Fig ~\ref{fig:eg1}(b) show that CF is similar to what many of us would expect in terms of conflict. For example, the algorithm gives higher conflict measure to $\{x_1, x_2, x_4\}$ than $\{x_1, x_2, x_3\}$. This is because source 3 has more overlap than source 4, to sources 1 and 2.

    Looking through the CF measures, we can find that the highest difference happens between $g^{CF}(\{x_1, x_2, x_3, x_4\})$ and $g^{CF}(\{x_2, x_3, x_4\})$ without considering the lowest layer. This shows that by adding source 1, the conflict measure rises the most, which means that source 1 can be identified as the most conflicting source.
   \end{example}

   \begin{example}
   In the second example, the four interval-valued evidences are $\overline h_1 = [10, 12]$, $\overline h_2 = [1, 4]$, $\overline h_3 = [1, 3]$, and $\overline h_4= [0, 2]$. The four evidences are shown in Fig. ~\ref{fig:eg2}(a), while the lattice of CF conflict measures is shown in Fig. ~\ref{fig:eg2}(b).

   \begin{figure}[t!]
    \begin{center}
    \subfigure[Four interval-valued sources with a unique source that has no overlapping region with others.]{\includegraphics[height=.09\textheight]{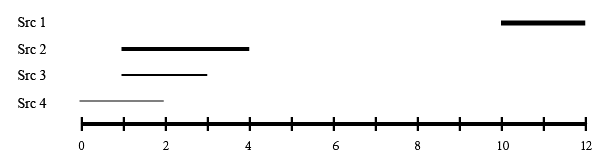}}
    \subfigure[Lattice of CF values.]{\includegraphics[height=.17\textheight]{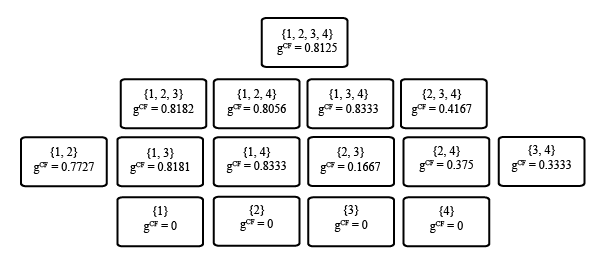}}
    \caption{Example 2: Moderate conflict.}
    \label{fig:eg2}
    \end{center}
    \end{figure}

    This example includes a unique source, i.e. source 1, which has no overlap with the other sources. Source 3 and 4 has the same length, and source 4 is further away from source 1 than source 3. Using CF, the measure value for $\{x_1, x_4\}$ is higher than the value for $\{x_1, x_3\}$, which is in accord with our expectation.
    The results in Fig. ~\ref{fig:eg2}(b) show that CF can well present the conflict among sources when there is unique source presented.

    Similar to Example 1, the highest measure value difference also happens between $g^{CF}(\{x_1, x_2, x_3, x_4\})$ and $g^{CF}(\{x_2, x_3, x_4\})$ without considering the lowest layer. This difference help us to identify source 1 as the unique (i.e. conflicting) evidence source.
   \end{example}

   \begin{example}
   In the third example, there are four interval-valued evidences, $\overline h_1 = [10, 12]$, $\overline h_2 = [4, 7]$, $\overline h_3 = [2, 4]$, and $\overline h_4= [0, 2]$. The four evidences are shown in Fig. ~\ref{fig:eg3}(a), while the lattice of CF conflict measures is shown in Fig. ~\ref{fig:eg3}(b).

   \begin{figure}[t!]
    \begin{center}
    \subfigure[Four interval-valued sources, and each source has no overlapping region with others.]{\includegraphics[height=.1\textheight]{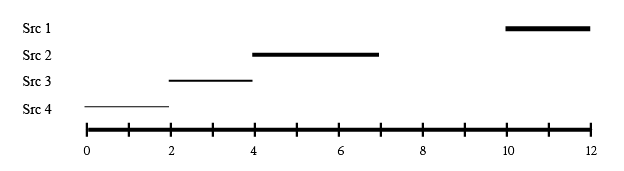}}
    \subfigure[Lattice of CF values.]{\includegraphics[height=.17\textheight]{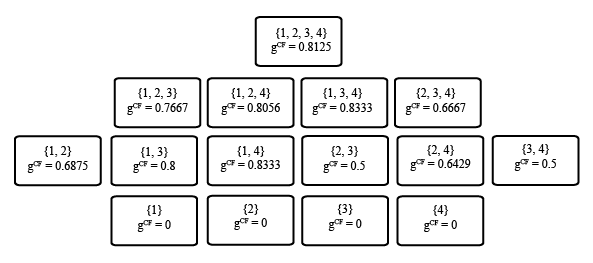}}
    \caption{Example 3: Extreme conflict.}
    \label{fig:eg3}
    \end{center}
    \end{figure}

  In this example, all four interval-valued evidences have no overlap with each other, which means this scenario is highly conflicted. It can be seen that all CF values in the upper three layers are higher than 0.5 and reach up to 0.8333. Comparing the three examples, every corresponding measure value in Example 1 and Example 2 is lower than or equal to the value in Example 3.
  \end{example}

   \subsection{Experimental Example}
\begin{figure}[t!]
    \begin{center}
    \subfigure[Measurement from four temperature sensors. ]{\includegraphics[height=.22\textheight]{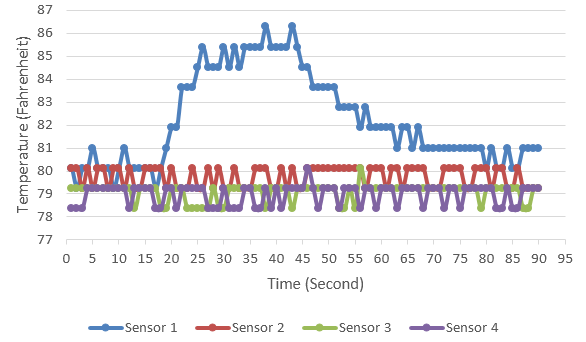}}
    \subfigure[Measure values of $g^{CF}(\{x_1, x_2, x_3, x_4\})$.]{\includegraphics[height=.22\textheight]{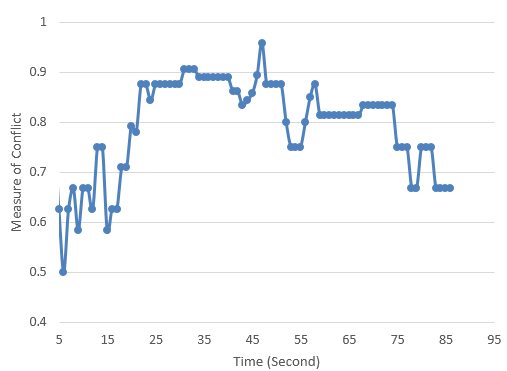}}
    \subfigure[Measure values of $g^{CF}(\{x_2, x_3, x_4\})$.]{\includegraphics[height=.22\textheight]{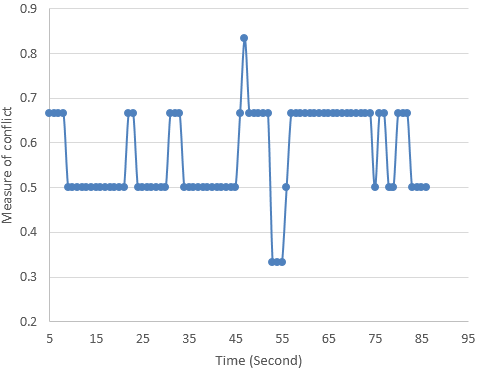}}
    \caption{Experimental example: (a) measurement from four temperature sensors, (b) CF among four sensors, and (c) CF among three sensors.}
    \label{fig:eg4}
    \end{center}
    \end{figure}

   \begin{example}
   In this section, we use data collected from four temperature sensors as our sources. As shown in Fig. ~\ref{fig:eg3}(a), the sensors collect data once per second for 90 seconds. During this period, the temperature detected by sensor 1 rises due to an external influence. We select five seconds for our time interval as a sliding window, and choose the maximum and minimum temperatures during the five-second period as the upper and lower endpoints for the intervals for each sensor. The five-second interval is chosen for convenience, and other choices would work as well. In Fig. ~\ref{fig:eg4}(b), the beginning of x axis is 5, which means that the related conflict measure is calculated using the period from 0 to 5 seconds. The measure used in Fig. ~\ref{fig:eg4}(b) is $g^{CF}(\{x_1, x_2, x_3, x_4\})$.

   It can be seen that as the first temperature sensor output rises, the CF value also rises. It shows that values calculated using CF can be used as an indication of changes among sources.

   In Fig. ~\ref{fig:eg4}(c), the measure used is $g^{CF}(\{x_2, x_3, x_4\})$, which is more stable than the values shown in Fig. ~\ref{fig:eg4}(b). This shows that sensor 1 is the most conflicting source among the four sensors.
   \end{example}

\section{Conclusions and future work}\label{sec:conclusion}
In this paper, we develop a measure of conflict in interval-valued settings as a normal measure.
The results shows that the output from CF has similar meaning to our ``common sense'', and it can help us identify conflicting sources.

In the future, we will present more algorithms that measure conflict, and compare them comprehensively.
We will develop more formal procedures to identify conflict in the resultant NM.
We will also extend the algorithm to set-valued information (e.g., probability distributions versus interval-valued information).
Last, we will do more experiments in real-world applications using sources, such as cameras, depth/range sensors or electrocardiograms.

\bibliographystyle{IEEEtran} 
\bibliography{IEEEfull,Refs}


\end{document}